\documentclass[twocolumn,10pt]{article}

\usepackage[a4paper,margin=0.5in]{geometry}
\usepackage{authblk}
\usepackage{graphicx}
\usepackage{amsmath,amssymb}
\usepackage{cite}
\usepackage{hyperref}
\usepackage{abstract}
\usepackage{caption}
\usepackage{titlesec}
\usepackage{comment}
\usepackage{array}
\usepackage{booktabs}

\titlespacing{\section}{0pt}{12pt}{6pt}

\title{\bfseries Zone-sectored organic crystals with spatially resolved exciton dynamics}

\author[1,2]{Moha Naeimi}
\author[1,2]{Tim Völzer}
\author[1,2]{Regina Lange}
\author[2,3]{Kevin Oldenburg}
\author[1,2]{Stefan Lochbrunner}
\author[1,2]{Ingo Barke}
\author[1,2]{Sylvia Speller}
\affil[1]{\small Institute of Physics, Universit\"at Rostock, 18059, Rostock, Germany}
\affil[2]{\small Department Life, Light \& Matter, Universit\"at Rostock, 18059 Rostock, Germany}
\affil[3]{\small Center for interdisciplinary electron microscopy (ELMI-MV), Department Life, Light \& Matter, Universit\"at Rostock, 18059 Rostock, Germany}
\date{}

\begin{document}
\twocolumn[
\maketitle
\vspace{-2em}
\begin{onecolabstract}
\noindent
Among the organic semiconductors, rubrene stands out in terms of hole mobility, luminescence yield and exciton migration distance. A novel type of rubrene microcrystal is prepared in the orthorhombic phase, exhibiting zone-sectored tabular domains with distinct photoluminescence (PL) characteristics. These sectors exhibit distinct PL spectra and time-evolution, arising from differences in the in-plane orientation of the orthorhombic unit cell relative to the crystal surface. A combination of polarised optical microscopy, fluorescence lifetime imaging microscopy (FLIM), and atomic force microscopy (AFM) is used to characterise the samples in terms of crystal orientation, fluorescence lifetime, and photoluminescence spectra. Spatially resolved PL spectroscopy reveals that the redshifted 650 nm emission band has polarisation along the transition dipole moment and is associated with high photon absorption due to the alignment of excitation polarisation and transition dipole moment and selectively localized within specific sectors of the crystal. The detected photon originates from direct emission of a geminate coherent triplet pair, or from its fusion. This band exhibits pure mono-exponential dynamics with 3.7 ns lifetime. The triplet fusion behaviour in the succeeding time regimes can be treated in the framework of power law scaling and random walk. 
The emission kinetics are modelled using rate equations describing geminate and non-geminate exciton fusion processes, enabling a quantitative interpretation of the spatially resolved PL kinetics. These findings introduce a material-based strategy, opening novel routes for photonic applications and light harvesting.

Keywords: rubrene, FLIM, AFM, triplet exciton migration, triplet exciton fusion
\end{onecolabstract}
\vspace{1em}
]
\section{Introduction}
Supramolecular dye crystals hold promise of efficient energy and information transfer via long-distance transport of excitons.  
Excitons are mobile quasiparticles made up by bound electron-hole pairs. They carry no net electric charge, and, in the singlet configuration, no net spin. They may be considered as an alternative carrier type with less Joule heating. Particularly, triplet excitons emerging from swift singlet exciton fission can propagate far~\cite{yost2012}.

Crystalline rubrene structures have been subject of intense research in the context of organic electronics \cite{McGarry2013}, e.g., for field-effect transistors \cite{Kim2007} and organic light-emitting diodes \cite{Wang2023}. Rubrene is a small organic molecule that can be considered a tetracene derivative with four additional phenyl rings. It exhibits several notable electronic properties, including high charge mobility \cite{vanderLee2022, Zhang2010, Nitta2019}, efficient singlet fission resulting in triplet excitons with a long lifetime and long diffusion lengths \cite{Wu2021, Breen2017, Finton2019}, triplet-triplet annihilation \cite{Baronas2022, Engmann2019}, and triplet energy transport \cite{Baronas2022}. These properties make rubrene a promising candidate for anisotropic energy and exciton transfer. 

One of the striking features of rubrene and of unsubstituted tetracene is, that triplet-pair (T$_1$T$_1$) and the singlet (S$_1$) state are energetically almost resonant, so that singlet fission and geminate triplet-triplet fusion occur as dominating processes. 
Quantum beatings have been observed for rubrene crystals, however with the help of an external magnetic field \cite{Wolf2018}, which changes the crystal field splitting parameter. For unsubstituted tetracene, the beatings become stronger with crystallinity, and with temperature \cite{Bardeen2012}.

It is now well known that singlet fission in rubrene is more efficient in crystalline phases \cite{Ma2012} while it is significantly weaker in amorphous structures \cite{Chen2020}. Hence, crystalline structures are necessary to provide a platform for extracting rubrene's unique opto-electronic properties. 

Additionally, the photoluminescence spectra and lifetime characteristics of rubrene, with their unique signatures, provide valuable insights into its optical and charge transfer properties.

Beyond the luminescence peak at ~550 nm, which corresponds to the radiative transition from the LUMO to the HOMO, two distinct and broad peaks are consistently reported at ~600 nm and ~650 nm \cite{Podzorov2011, Wolf2021,Ni2021,Ma2013}.

While the ~600 nm peak is dominant in rubrene single crystals for excitation polarisations in the \textit{ab} plane, Wolf et al. \cite{Wolf2021} note that the luminescence band at 650 nm decays within a time window below 50 ns, possibly indicating short-range non-geminate exciton-exciton annihilation. 

Macroscopic crystals of rubrene can be grown using the physical vapour transfer (PVT) method \cite{Zeng2007, Laudise1998}, which has enabled fascinating experiments. With this method, large and high-quality crystals are grown by placing source material with relatively high vapour pressure inside a container and heating it to the vapour phase. To ensure a high crystal quality, the vapour is guided through a tube with a controlled atmosphere to a cooler environment where the condensation seeds the crystals on the substrate. 

However, preparing extended rubrene crystals directly on a substrate is challenging because rubrene tends to become trapped in a metastable state of low order or an unfavourable crystal structure during the preparation process.
The challenge essentially lies in the fact that providing sufficient thermal energy to overcome this metastable state easily results in massive desorption, eventually evaporating all available material from the substrate.

A variety of methods have been proposed, which are based on temperature treatments or epitaxial methods \cite{Chang2015, Wei2022, AkinKara2022, Luo2007, Park2007}. Regarding temperature treatment, the main idea is to reach the crystallisation temperature while preventing desorption, usually by providing a high partial pressure of rubrene in the local environment of the substrate.

In the pursuit of preparing compact, stable, yet thin and flat single crystals, we report a growth method that results in the formation of a new type of tabular single microcrystal exhibiting distinct sectors.
In this work we explore the optical properties and exciton dynamics of these highly crystalline molecular assemblies by examining their structure, morphology and luminescence. We refine our method to control the growth of rubrene crystalline structures by combining abrupt-heating \cite{Lee2011} and microspacing sublimation techniques \cite{Ye2018}. We have recently reported on the growth habits of rubrene on highly oriented pyrolytic graphite (HOPG) as substrate \cite{RubHOPG}, demonstrating that pre-heated rubrene coverage is a critical factor influencing the crystal growth. Throughout this study, no co-materials, such as buffer layers, are used to facilitate growth \cite{Verreet2013, Tan2023, Fusella2017, Foggiatto2019}.

\section{Experimental section}
Rubrene powder with a purity of 99.999~\% (Sigma-Aldrich) was used as received and the container was kept in a desiccator to prevent oxidation. The steps of sample preparation and resulting crystals will be discussed in the next section.

After preparation, the crystals were investigated using polarisation optical microscopy (Zeiss Axio lab 5) with a colour camera (Axiocam 305 colour R2) and a bright light LED 10W as the light source. For fluorescence microscopy an Olympus IX73 inverted fluorescence microscope was used.

The fluorescence setup employed two 100 nm wide bandpass filters with centre wavelengths at 470 nm and 632 nm, and a longpass beamsplitter with a cutoff wavelength of 550 nm. Surface morphology investigations were conducted  using atomic force microscopes (Park Systems NX20 and XE100) equipped with conductive tips made of chromium platinum (Cr-Pt) exhibiting a cantilever spring constant of 3 N/m and a free eigenfrequency of 75 kHz.

Photoluminescence spectra and time-resolved measurements were obtained with a confocal setup using a fluorescence lifetime imaging microscope (MicroTime 200, PicoQuant).
The sample was excited by laser pulses at 532 nm with various pulse energies at a repetition rate of 20 kHz. The laser was focused to a spot with a diameter of $1~\mu m$. Luminescence light was detected by a time-correlated single photon counting detector (Kymera 193i).

Visualisation and data analysis were performed using Gwyddion \cite{Gwyddion}and Igor Pro (Wavemetrics).

\section{Growth setup}

Using the methods of microspacing sublimation \cite{Ye2018} and abrupt heating \cite{Lee2011}, we controlled the growth of rubrene single crystals, aiming for the orthorhombic crystal phase with four key preconditions (Figure S1). 

First, rubrene powder is deposited on a substrate (acting as source). The deposition of preheated rubrene onto the substrate must be carefully controlled. In our previous work \cite{RubHOPG}, we demonstrated that a superfine coverage of preheated amorphous rubrene on the substrate promotes the formation of the desired orthorhombic phase. Second, amorphous rubrene is subjected to heating at temperatures significantly higher than its sublimation point, so that the crystallisation starts before desorption. Third, to facilitate crystallisation over desorption at high annealing temperatures, a confined growth environment with elevated partial pressure should be maintained \cite{Ye2018}. Finally, thermal energy transfer to amorphous rubrene must be rapid to prevent it from becoming trapped in metastable phases at the second and third crystallisation temperatures, 140°C and 170°C, respectively \cite{Fielitz2016}.

\begin{figure*}
    \centering
    \includegraphics[width=1\linewidth]{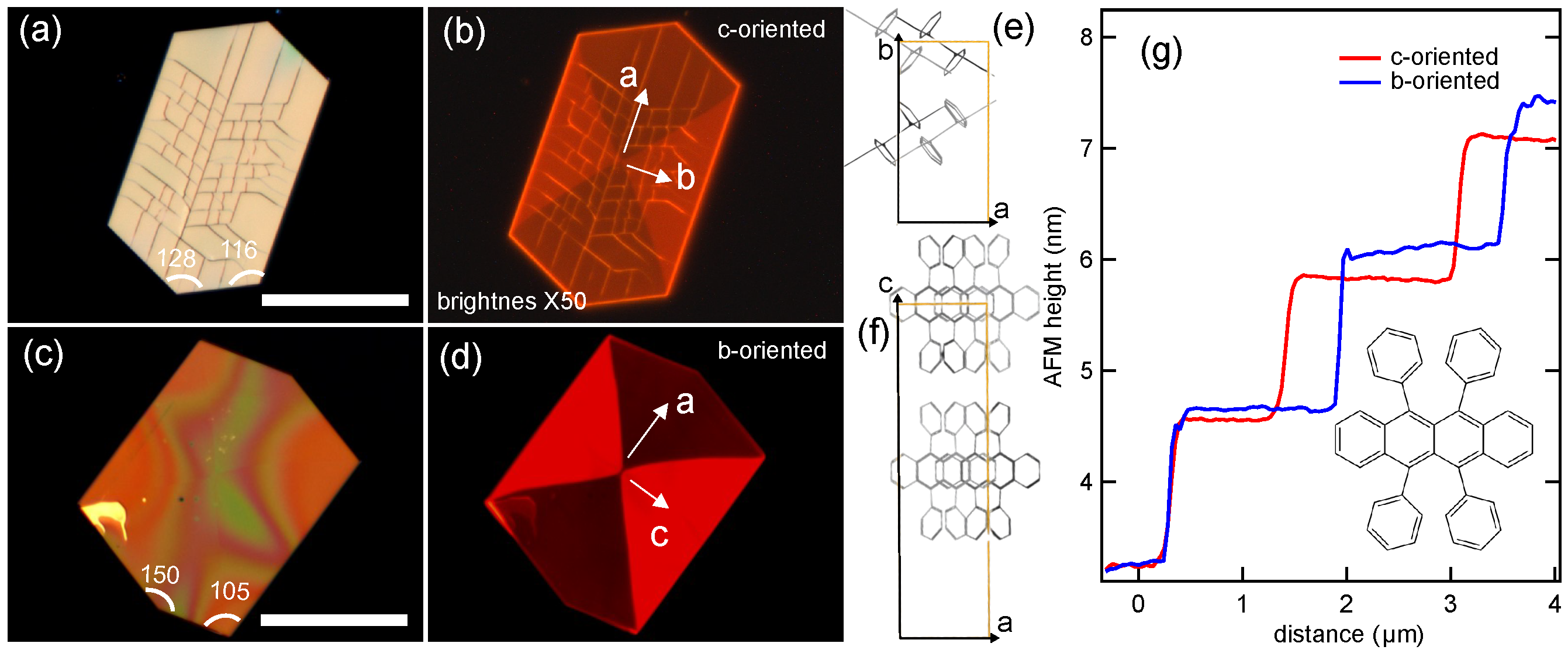}
    \caption{Two different crystal types prepared by the presented method. (a) and (b) POM and fluorescence images of a single rubrene crystal having an orthorhombic phase with \textit{c}-axis as the out-of-plane axis. (c) and (d) POM and fluorescence images of a single rubrene crystal grown along the \textit{b}-axis of the orthorhombic unit cell. (e) Stacking of the rubrene molecules in the \textit{ab}-plane and (f) \textit{ac}-plane of the orthorhombic unit cell. The yellow rectangles represent the schematic lattice parameters of the orthorhombic unit cell having the lengths of \textit{a}=0.718 nm,\textit{b}=1.44 nm and \textit{c}=2.69 nm. (g) AFM height profiles from high resolution topography of the two different types. Inset: rubrene molecule. }
    \label{fig:hourglass}
\end{figure*}

To achieve these four preconditions, we lightly abraded a small amount of rubrene powder onto a source holder (silicon substrate covered with a native SiO$_2$ layer) using a weak force applied by a clean spoon in a clean, well-isolated environment to maintain purity.

Through abrasion, the large grains within the powder crumble, achieving a more uniform distribution on the source holder. Importantly, this process does not involve rearrangements on a molecular scale, contrasting with methods like dissolving rubrene in a solvent or evaporating it. Another advantage of abrading is the reduction of air inclusions in the source powder and airborne hydrocarbon species. These species are even able to fully replace the primary molecule \cite{Plinks2022, Cholakova2019}.
Therefore, reducing air inclusions in the source powder --- a consequence of rubbing --- may be crucial (see Figure S2 for impact of pre-treatment of rubrene).

Another SiO$_2$ substrate was placed approximately 150 µm above the source holder, using tantalum slides as spacers, with its surface facing down toward the source holder. The choice of the spacer material ensures effective high-temperature transfer between both substrates. We avoided evaporation or spin-coating to prevent disruption of the intermolecular arrangement \cite{Guo2020} and impurities. 

The assembly was then placed onto a preheated hot plate. The temperature of the sample was carefully controlled and measured with a K-type thermocouple to anneal the amorphous phase at a temperature well above the second crystallisation stage and the rubrene sublimation temperature of 270°C for 10 minutes.

\section{Results and discussion}
Figures \ref{fig:hourglass}a and \ref{fig:hourglass}b show cross-polarised light (POM) and fluorescence microscopy images of a rubrene single crystal with a low luminescence yield prepared by the method described above. The fluorescence images are taken with unpolarized excitation- and detection- polarisation. Selected area electron diffraction (SAED) shows diffraction spots and confirms crystallinity (figure S6). However, it turns to be cumbersome to identify the crystal structure. Hindrances were tearing during transfer to the XRD tip and melting due to electron irradiation, on a TEM grid. Furthermore, we rely on the crystal habit, which suggests, according to Massimo Moret et.al \cite{Moret2022}, a growth shape and orthorhombic structure (QQQCIG Cambridge structural data base). 
The crystal exhibits growth shapes \cite{Moret2022} consisting of dark diamond- and bright triangular-shaped zones, which are not visible in the POM contrast.
Figure \ref{fig:hourglass}c and \ref{fig:hourglass}d shows another rubrene crystal prepared by the same method, exhibiting high luminescence yield. Also in this case dark diamond and bright triangular zones can be recognized, with a pronounced contrast.

The fringe patterns in Figure \ref{fig:hourglass}a with different colours are due to different local thickness of the crystal, as confirmed by topography inspection (by AFM). The weakly fluorescing crystal, however, exhibits line defects (cracks) on its surface.

These crystals featuring an ”hourglass” signature \cite{Kahr2004, Vetter2002}, which is reported for the first time for rubrene and reminiscent to embedded little hourglasses, appear to exhibit four different zones of two types, i.e., diamond and triangular.
This is a rather peculiar signature that has been very recently reported for perylene crystals, however, obtained with a different growth method \cite{Takazawa2024}. It is related to hourglass sector zoned minerals \cite{Beno2020}. In mineralogy, it is reported that the zones behave very differently in taking up trace species such as atoms or dye molecules, so that they can be made visible by colour or by energy-dispersive X-ray analyses in SEM or by cathodoluminescence. It is put forward that the lattice of the two types of zones does not differ.

It should be emphasised that the present method also results in the growth of long one-dimensional rod-shaped rubrene crystals and conventional microcrystals without the hourglass signature (see Figure S2).

The origin of the hourglass sectoring is known in mineralogy, however not fully clear\cite{Ubide2019}. In particular as the striking characteristic of well-defined internal diamond and triangular zones in fluorescence images is only observed for the presented preparation method.
Interestingly, all the crystals with high fluorescence yields adopt angles between the crystal edges of 150°and 105° (Figure \ref{fig:hourglass}c) or of  75°and 105° , while all of the crystals exhibiting angles of 128° and 116° (Figure \ref{fig:hourglass}a), as well as 120° (i.e.\ perfect hexagons) are of the weakly fluorescing type. The diamond and triangular zones are not visible in the surface morphology, hence the origin must be due to internal optical properties (Figures S4).

The hexagonal tabular crystals are fully compatible with the growth habit of the orthorhombic crystal phase of rubrene \cite{Moret2022} (Figure \ref{fig:hourglass}e and \ref{fig:hourglass}f). The growth habit is kinetically determined by anisotropic growth speeds while the equilibrium habit is the result of thermodynamic equilibration.

The fast growth direction is along the eight physically equivalent \textlangle 111\textrangle\ directions, giving rise to eight rectangular pyramids, touching each other with their tips at angles of 30° and 150° depending on whether they are opposed or adjacent. Their growth evolves from tip to base. The \textlangle 111\textrangle\ directions roughly
coincide with the a-direction of the orthorhombic lattice. The slower growth along the c and b axes takes place somewhat delayed. It may happen, that such later grown sections do not monolithically connect to the earlier pyramids and rotated domains emerge. Then, the transition dipole moments are also along different directions than in the early grown crystal sections. This becomes recognizable by optical methods applying defined polarisations, i.e. in case of linearly polarised laser excitation or polarisation sensitive luminescence. 

The striking differences in luminescence yield among various crystal types and domains are attributed to the orientation of the molecular plane, specifically the transition dipole moment. In the following sections, the two different types of crystals will be discussed based on their photoluminescence spectra and kinetics.

The molecular transition dipole moment of the S0-S1 transition is aligned along the short axis of the rubrene molecule (in the aromatic plane). In the unit cell, all molecules are oriented with their short axis parallel to the \textit{c}-axis and thus also the transition dipoles. In the crystals with low fluorescence yield the \textit{c}-axis and the transition dipoles are oriented perpendicular to the surface. 
Consequently, vertically impinging excitation light with in-plane polarisation is weakly absorbed, and singlet emission occurs primarily laterally.
As a result, these crystals appear dark with a bright rim, as has been shown earlier \cite{Irkhin2012}. We call this crystal type "\textit{c}-oriented" since the \textit{c}-axis of the orthorhombic unit cell is perpendicular to the crystal plane (see Figure \ref{fig:hourglass}f).

\begin{figure*}
    \centering
    \includegraphics[width=0.9\linewidth]{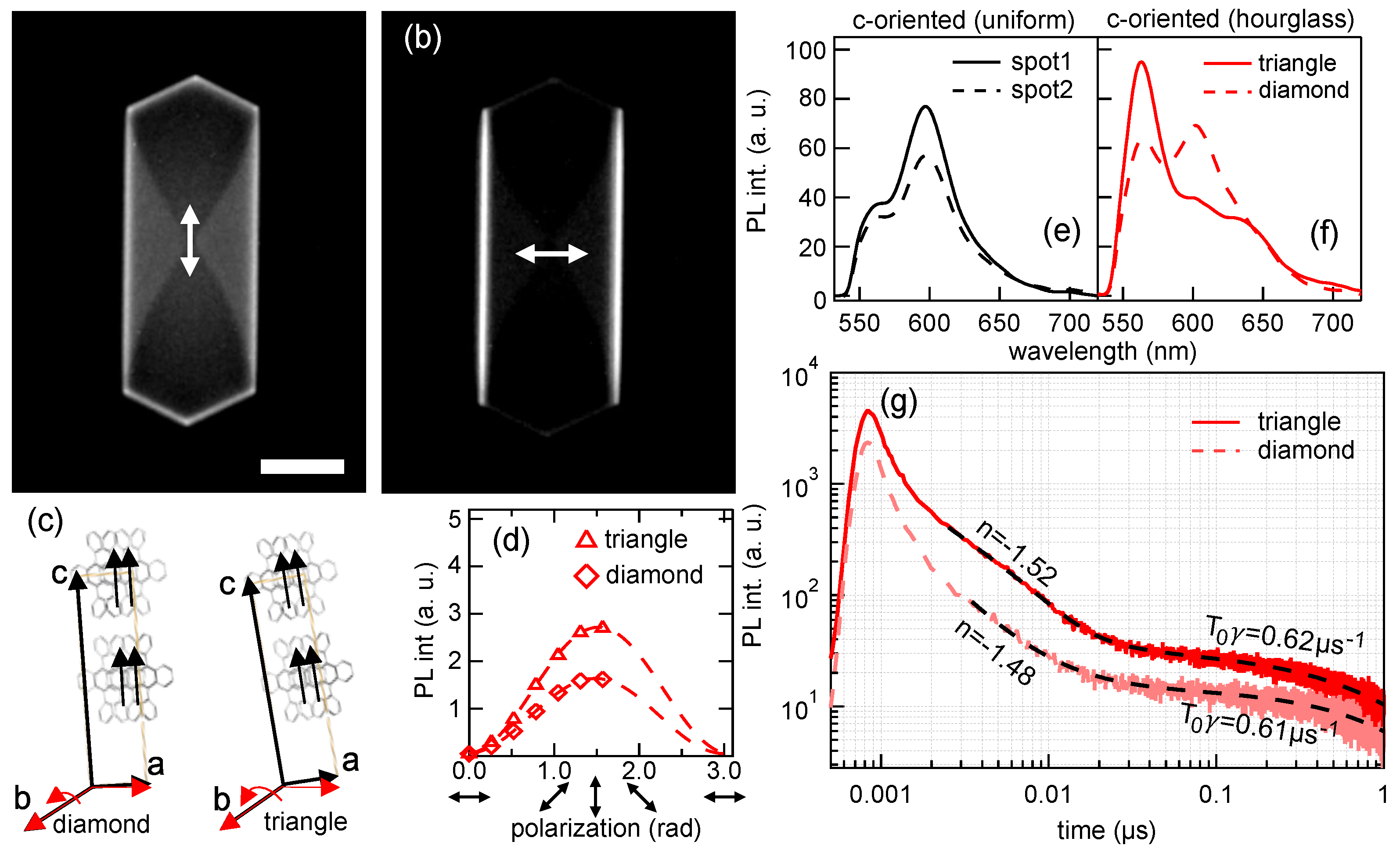}
    \caption{The origin of the contrast between diamond and triangular zones is a tilt of the lattice orientation. (a) and (b) Fluorescence images of \textit{c}-oriented rubrene single crystal with excitation polarisation along \textit{a}-axis and \textit{b}-axis, respectively. scale bar: 20 $\mu$m (c) Schematic representation of the tilt of the orthorhombic unit cell around the \textit{b}-axis. The red straight arrows represent the polarisation direction. (d) Fluorescence intensities versus polarisation of excitation light from $0^{\circ}$ (along \textit{b}-axis) to $90^{\circ}$ (along \textit{a}-axis). (e and f) luminescence spectra of a normal \textit{c}-oriented crystal and an hourglass \textit{c}-oriented crystal, respectively. (g) luminescence kinetics of an hourglass \textit{c}-oriented crystal resolved for diamond and triangular zones.}
    \label{fig:cor}
\end{figure*}

In the strongly fluorescing type, transition dipole moments of the S transition are oriented largely in plane, i.e. in the \textit{ac} plane, hence the absorption of photons with in-plane polarisation is possible and peaks when the excitation polarisation aligns with the \textit{c}-axis.
They show up bright in luminescence and we refer this type to as "\textit{b}-oriented", since the \textit{b}-axis is the out-of-plane axis. Note that the basis vectors \textit{a} (= 0.718 nm), \textit{b} (= 1.44 nm), and \textit{c} (= 2.69 nm) correspond to the shortest, intermediate, and longest vector of the orthorhombic unit cell of rubrene, respectively. 

The both crystal types exhibit slightly different step heights (Figure \ref{fig:hourglass}e). This corroborates the attributed orientations and both types belong to the orthorhombic crystalline phase, in line with \cite{ElHelou2010}.
We tentatively attribute the intra-crystal contrast—i.e., the diamond- and triangle-shaped zones seen in Figure \ref{fig:hourglass}—to crystal twinning, as these regions respond differently to polarisation. The exact nature of this twinned phase could involve a rotation of the orthorhombic unit cell around the planar axes (see below).

\subsection{Characteristics of the c-oriented type}
For the \textit{c}-oriented type, where the dipole moment is perpendicular to both the crystal surface and the polarisation plane, a slight polarisation dependency of emission is observed. Figures \ref{fig:cor}a and \ref{fig:cor}b show fluorescence images of a \textit{c}-oriented rubrene crystal under excitation light, polarised along the \textit{a}-axis and \textit{b}-axis, respectively. As the polarisation rotates, the photon yield varies periodically, reaching its maximum when the excitation is \textit{a}-polarised. Conversely, when the polarisation is along the \textit{b}-axis, emission is minimal. A similar trend is observed for emitted light, indicating that the emission follows the transition dipole orientation. This periodic behaviour is illustrated in Figure \ref{fig:cor}d.

Since the polarisation of the excitation light is always in-plane—i.e., perpendicular to the surface normal of the crystal—the observed variations in emission intensity in diamond and triangular sectors suggest a rotation of the unit cell. This rotation brings the dipole moment partially in-plane, those sectors exhibiting stronger emission corresponding to more rotated unit cells.

The excitation and emission from a side facet of \textit{c}-oriented crystal is also polarisation dependent. The (101) facets \cite{ElHelou2010} exhibit maximal intensity under \textit{b}-polarised light, as this orientation aligns the dipole moment with the excitation polarisation. Similarly, the (111) group reaches maximal intensity when the polarisation is oriented perpendicular to the rim of the facet.

An interesting observation is the variation in the photoluminescence spectra across different zones. Here, vertical illumination at 532 nm with in-plane polarisation relative to the sample surface is applied and the detection is unpolarised. Figure \ref{fig:cor}e shows the luminescence spectra from two different spots of a \textit{c}-oriented crystal with absent zone-sectoring—i.e., no diamonds or triangles. According to the literature, the band at 564 nm corresponds to the zero-phonon transition from the ground state to the first electronically excited singlet state of the molecule and is \textit{c}-polarised \cite{Ma2013}. It is noted that the intensity of this band is strongly reduced by self-absorption inside the crystal, since the red wing of the absorption spectrum overlaps with this emission band and is also c-polarised. The luminescence band at 600 nm is probably induced by a carbon-carbon stretch vibration breaking the C$_{2h}$-symmetry of the rubrene molecule \cite{Irkhin2012}. This results in a transition dipole along the b-axis. The measured spectra are then in line with the expectation that an excitation polarisation perpendicular to the \textit{c}-axis enhances the 600 nm emission.

When measuring the luminescence spectra of a \textit{c}-oriented hourglass crystal —i.e., containing diamonds and triangles (Figure \ref{fig:cor}f) the spectral features vary between the different zones. As discussed earlier, zone-sectoring originates from the rotation of the orthorhombic unit cell among these sectors. This rotation is more pronounced in the triangular sectors, leading to an increased in-plane component of the \textit{c}-axis that aligns with the excitation polarisation. Consequently, the 564 nm band becomes dominant in the triangular sector, while a faint 646 nm band also emerges. In contrast, in the diamond sectors—where the rotation of the unit cell around the b-axis is less pronounced—the 600 nm band remains dominant, and the 646 nm band is absent. This is compatible with our earlier interpretation concerning the rotation of the unit cell. We should mention however, that these rotations are subtle since both the photoluminescence intensity and spectra are roughly similar to those from a normal c-oriented crystal with no hourglass signature.

Another way to gain insight into the nature of the different parts of the crystal is by examining their emission lifetimes and kinetics. Figure \ref{fig:cor}g is a double logarithmic plot showing the photoluminesce kinetics of a \textit{c}-oriented hourglass crystal resolved for diamond and triangular sectors. The signal from the bright rims of the crystal is not included into the data. The exciton dynamics in \textit{c}-oriented crystals are largely similar for both sectors; however, the subtle differences are interesting. For both regions, there is a turning point, suggesting different exciton processes before and after 20 ns.

Here we interpret the kinetics with two distinct processes, i.e. geminate and non-geminate fusion. Directly after the fission of a singlet state into two triplet excitations geminate fusion can occur \cite{Wolf2021}. In this process the two triplet excitations recombine to form again the singlet state. The latter can then emit during its short lifetime a fluorescence photon or, much more likely, undergo again fission and dissociate into two triplets. If one of the triplets separates from the other by a hopping step fusion is no longer possible for the time being. However, the triplets perform random hops along the crystal axis and can accidentally encounter again resulting in the chance of a further fusion event. Already in previous work on crystalline rubrene and tetracene, geminate fusion of the triplet excitons was studied \cite{Wolf2021, Seki2018, Seki2021, Bulovic2014}. It was shown that this process leads to delayed fluorescence which decays approximately with a power-low, i.e. the signal is $\propto t^n$ and the exponent n is negative. Interestingly, it turns out that n = -1.5 in the case of three-dimensional hopping and n = -1 in the case of two-dimensional \cite{Merschjann2015}. In case of resonant exciton fission-fusion materials, the situation is special: One can assess diffusion not merely via a spatial observable, but also from temporal information. In that case, the dimensionality is no longer in the prefactor, but in the exponent \cite{Pfluegl1998,Seki2021}. This is the key to extract from a non-spatial information, in our case time-dependent fusion events, on spatial diffusion. The power law fit turns out to have a similar exponent of $\sim$ -1.5 for both sectors, indicating diffusion in three dimensions.

This geminate process is then followed by non-geminate fusion. This contribution describes the time dependence of the emission caused by non-geminate fusion of diffusing triplet excitons. Assuming the fusion dynamics can be characterised by a bimolecular rate $\gamma$, the decay of the triplet concentration T obeys the rate equation $\frac{dT}{dt} = -\gamma T^2$ resulting in $T(t) = T_0(1+T_0\gamma)^{-1}$ \cite{Ryasnyanskiy2011}. $T_0$ is the initial concentration of the triplet excitons after singlet fission is completed. The singlet emission is proportional to the number of recombination events and thereby $\propto (1+T_0\gamma)^{-2}$. The fit described below yields a value for the product $T_0\gamma$ of the initial triplet concentration and the bimolecular rate. 

The following function is then suggested to quantify the PL kinetics:

\begin{equation}
\text{Signal} = A_1t^{n} + A_2(\frac{1}{1+T_0\gamma})^{2} +  \text{offset} 
\end{equation}

The results of the individual fits are given in Table \ref{tabel} and the obtained fit curves are shown in Figure \ref{fig:cor}g as black broken lines. Here $T_0 \gamma = 0.6  \mu s^{-1}$ in both diamond and triangular sectors is derived. For higher laser intensities, we record $T_0 \gamma$ up to $4 \mu s^{-1}$, while the power law exponent remains the same (Figure S3).

An interesting outcome from the fit is that a substantial background is required for a decent accuracy. Such background hints to the presence of emitting species with longer (life)times, i.e., involving longer migration paths or trap states. We cannot fully quantify the long lifetimes, since our experimental time window is limited to 1 $\mu$ s. But similar to Burdett et. al. \cite{Bardeen2010} one may expect a regime of 0.1 to few microseconds. Since our time window does not exceed 1 $\mu$s, the triplet lifetime of $> 100 \mu s$ is too long to be considered in our analysis. Such long lifetimes may occur upon late non-geminate fusion possibly in connection with transient self-trapping. In the fits a small offset in the range of a few counts has to be included (See Equation \ref{eq2}). It is probably related to long-lived species. It also cannot be excluded that self-trapped excitons or defects contribute to the long-lived signal \cite{Ni2021}.

\subsection{Characteristics of the b-oriented type}

\begin{figure}
    \centering
    \includegraphics[width=1\linewidth]{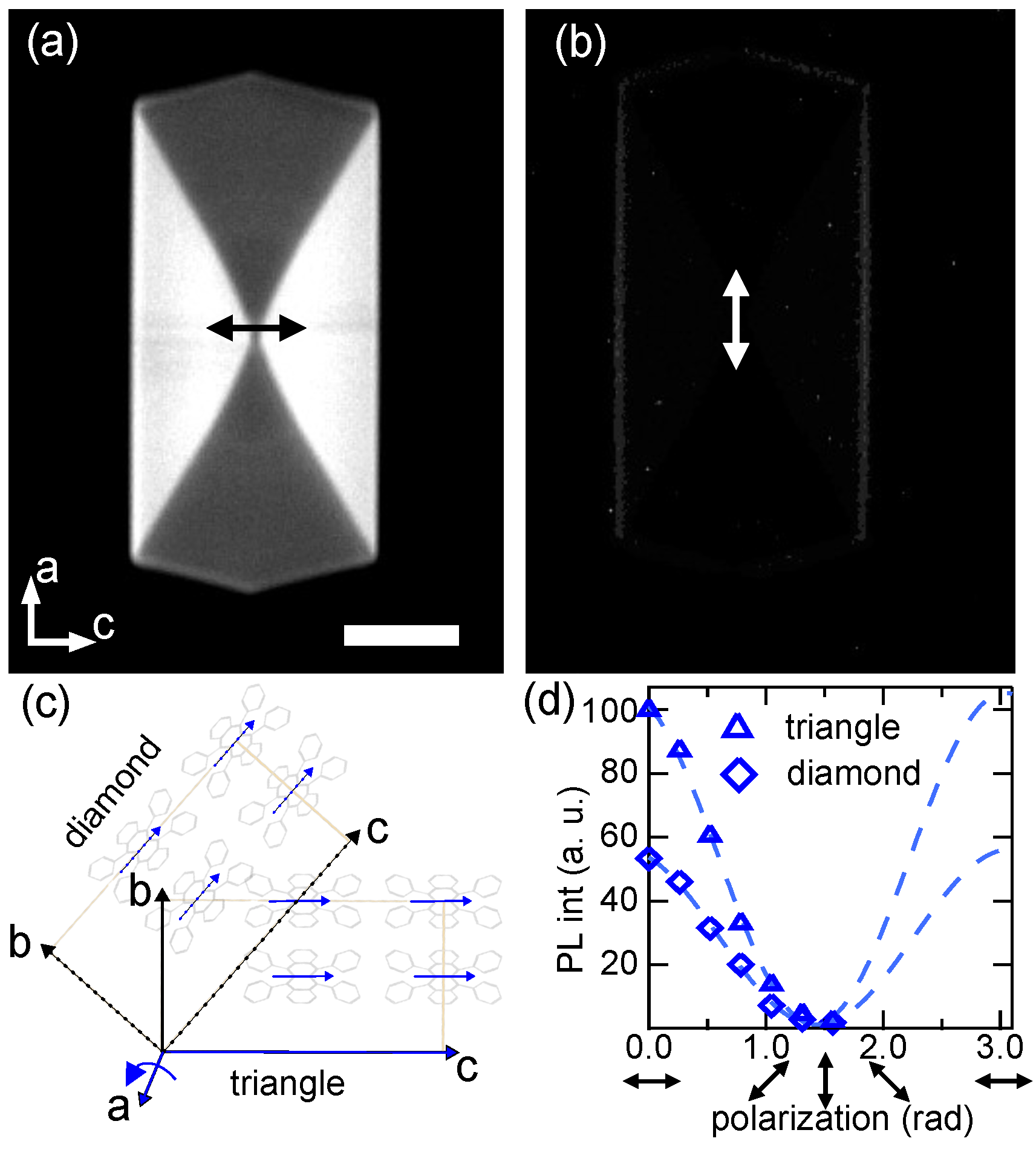}
    \caption{(a and b) Fluorescence images of a \textit{b}-oriented crystal with excitation polarisations along the \textit{c}- and \textit{a}- axes respectively. (scale: 20 $\mu$m) (c) Schematic representation of tilt of orthorhombic unit cell around the \textit{a}-axis in diamond sectors. The straight blue arrows represent the polarisation direction. (d) luminescence intensity versus polarisation direction from $0^{\circ}$ (along \textit{c}-axis to $90^{\circ}$ (\textit{a}-axis.)}
    \label{fig:bor}
\end{figure}

For the \textit{b}-oriented type, the transition dipole moments of the rubrene molecules are in the plane of the crystal surface and in the plane of excitation polarisation. Hence a pronounced emission and polarisation dependency is expected. Figure \ref{fig:bor}a and \ref{fig:bor}b show the fluorescence images of a \textit{b}-oriented crystal excited with \textit{c}-polarised and \textit{a}-polarised light, respectively.

When the excitation is \textit{c}-polarised, the fluorescence is in general very strong (Figure \ref{fig:bor}a). However, in the diamond zones it is approximately half that of the triangular zones, whereas both exhibit an equally minimal yield under \textit{a}-polarised excitation (Figure \ref{fig:bor}b). Figure \ref{fig:bor}d shows the PL intensity of the crystal shown in Figure \ref{fig:bor}a and \ref{fig:bor}b versus excitation polarisation. The weaker luminescence in the diamond zones suggests that the unit cell in these zones is rotated along the \textit{a}-axis versus those in the triangular zones, as the rubrene transition dipole moments then have partial instead of complete in-plane components (Figure \ref{fig:bor}c). The rotation angle in the diamond zones can be estimated using $\theta = \arccos(\sqrt{(\frac{I_{Dmax}}{I_{Tmax}}})$, where $\theta$ is the angle between the polarisation and transition dipole moment in the diamond sectors and $I_{Tmax}$ and $I_{Dmax}$ are maximum fluorescence yields in the triangular and diamond zones, respectively. We obtain varying angles ranging from 12$^\circ$ to 45$^\circ$ for different samples.

\begin{figure}
    \centering
    \includegraphics[width=1\linewidth]{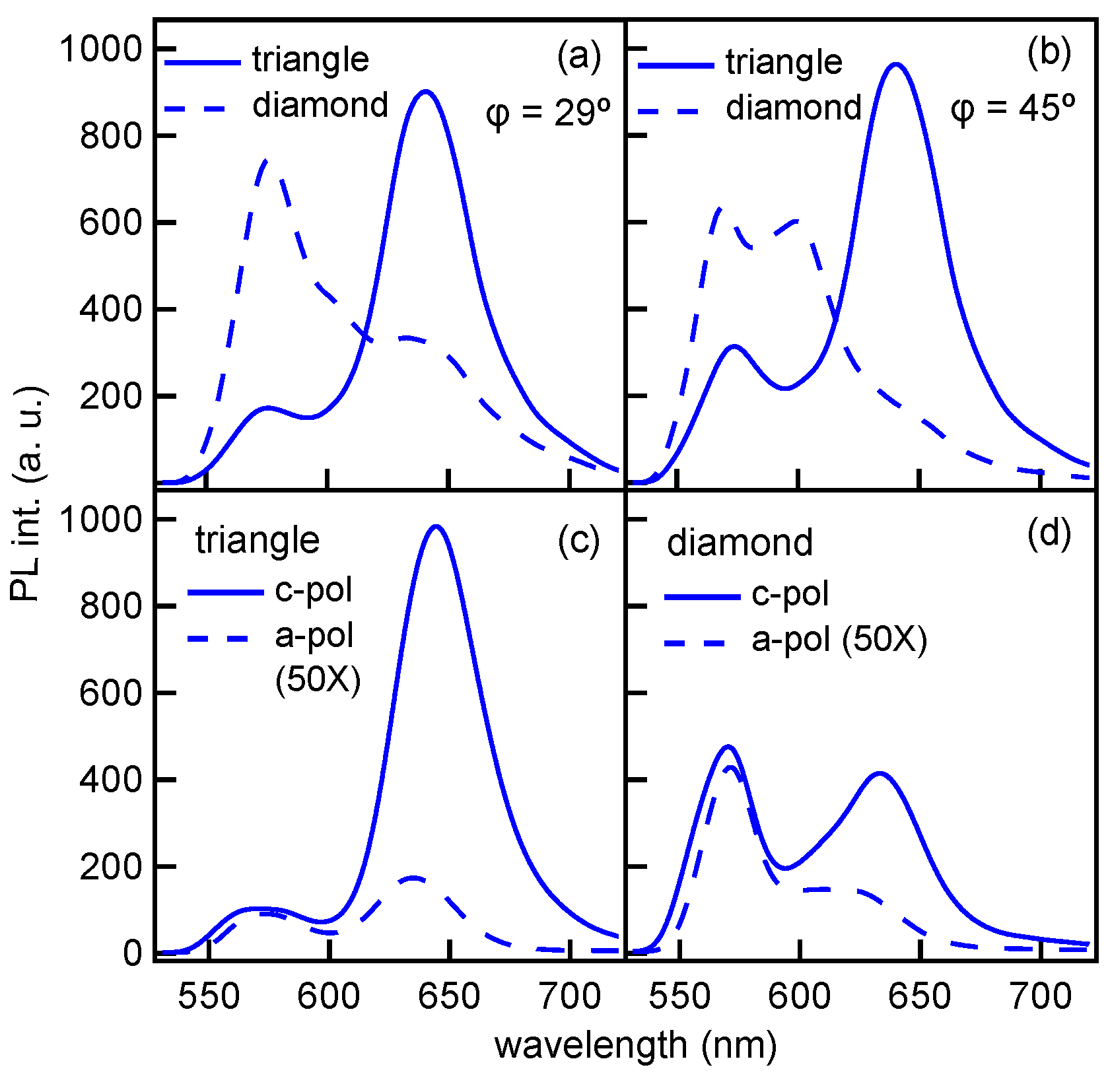}
    \caption{(a and b) Luminescence spectra taken from two \textit{b}-oriented crystals, resolved for different sectors, where the orthorhombic unit cell in diamond is rotated along the \textit{a}-axis by $29^{\circ}$ and $45^{\circ}$, respectively. The excitation and detection are unpolarised. (c and d) Luminescence spectra taken from the triangular and diamond zones, respectively. The excitation polarisation is along \textit{c}-axis and detection polarisation along \textit{c}- and \textit{a}- axes. For the detection along the \textit{a}-axis, the acquisition time is 50 times longer than along the \textit{c}-axis, in order to account for the low intensity and collect decent statistics.}
    \label{fig:spec_bor}
\end{figure}

Figures \ref{fig:spec_bor}a and \ref{fig:spec_bor}b show luminescence spectra of two different \textit{b}-oriented crystals, where in the diamond zones, the unit cell is rotated by 29$^\circ$ and 45$^\circ$ around the \textit{a}-axis, respectively. The spectra are characterised by three bands centred at 564, 600 and 646 nm whose amplitudes depend on the considered sector and the detected polarisation. As discussed earlier, the 600 nm band is probably induced by a vibration breaking the molecular C$_{2h}$ symmetry, resulting in a transition dipole along the \textbf{b}-axis. In the diamond sectors the rotation around the \textit{a}-axis results in a projection of this transition dipole onto the detected polarisation axis. Accordingly, the 600 nm band gains intensity with increasing rotation angle $\theta$ (Compare Figure \ref{fig:spec_bor}a and \ref{fig:spec_bor}b). Transitions due to the molecular vibronic progression of the molecule might contribute in the red wings of the main bands \cite{Irkhin2012}.

The triangular sectors seem to be strictly \textit{b}-oriented and show no 600 nm band, since the detection of \textit{b}-polarised emission is in this case impossible. However, their spectra are dominated by the 646 nm band, which is commonly referred to as the ”anomalous” band \cite{Wolf2021, Bossanyi2024}. Interestingly, the 564 nm band is weaker in these sectors, maybe due to stronger self-absorption resulting from the perfect alignment of the \textit{c}-oriented absorption dipole. In the diamond-shaped zones, the 646 nm band loses intensity with increasing rotation angle $\theta$. This results probably from the fact that the rotation reduces the projection of a \textit{c}-oriented transition dipole onto the detected polarisation.

Figures \ref{fig:spec_bor}c and \ref{fig:spec_bor}d show the luminescence spectra resolved for different zones, excited with polarisation along the \textit{c}-axis and detected parallel to the \textit{c}- and \textit{a}-axes, respectively. In the case of an \textit{a}-polarised detection a fifty times longer integration time was necessary to obtain a number of counts comparable to \textit{c}-polarised detection showing that the emission is strongly \textit{c}-polarised. Interestingly, the polarisation contrast of the 646 nm band is even stronger than that of the 564 nm band. 

\begin{figure}[h]
    \centering
    \includegraphics[width=1\linewidth]{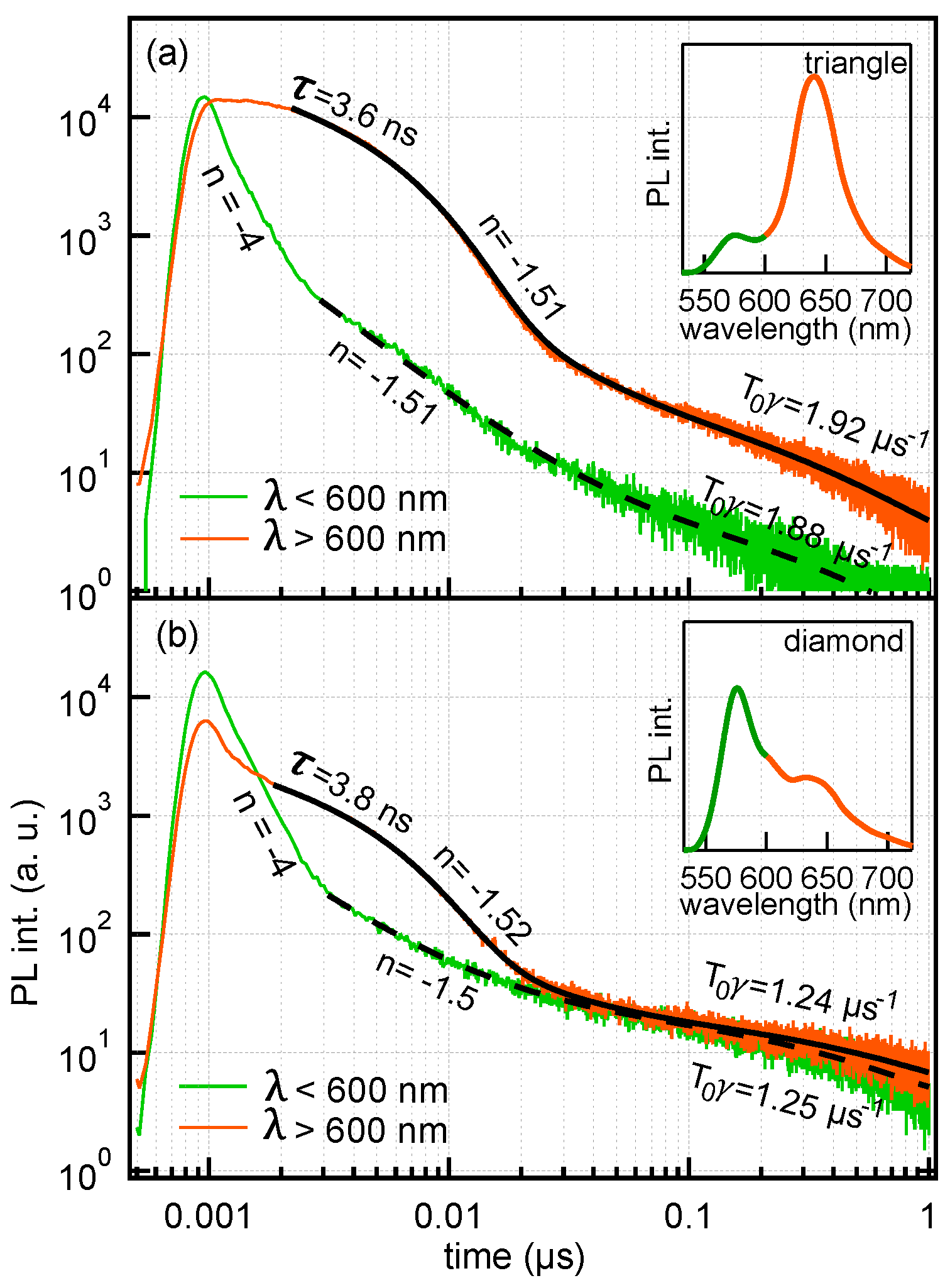}
    \caption{(a) Spectrally resolved luminescence kinetics of the triangular zone from a \textit{b}-oriented crystal. (b) Spectrally resolved luminescence kinetics of the diamond zone from a \textit{b}-oriented crystal. The inset is the respective luminescence spectra. }
    \label{fig:lt_bor}
\end{figure}

To gain further insight into the origin of this spectral band, we investigated the photoluminescence kinetics of each sector and each individual spectral band. Figure \ref{fig:lt_bor}a and \ref{fig:lt_bor}b show the photoluminescence kinetics of a \textit{b}-oriented crystal. We separate the kinetics into two different spectral regimes i.e. $\lambda < $ 600 nm and $\lambda >$ 600 nm, for triangular (\ref{fig:lt_bor}a) and diamond (Figure \ref{fig:lt_bor}b) sectors.

The kinetics of the emission with $\lambda > $ 600 nm, which is mostly related to the 646 nm band, is for the first 10 ns dominated by a mono-exponential decay, both in the triangular and in the diamond zones. The time constant of this process is $\tau \sim $ 3.8 ns. It indicates that the 646 nm band is due to emission from an electronically excited state with this specific lifetime. For times larger than 10 ns the kinetics are similar to the ones obtained for $\lambda < $ 600 nm.

For the latter spectral region, i.e. $\lambda < $  600 nm, the emission kinetics of the triangular as well as of the diamond sectors are strongly non-exponential and similar to those of the \textit{c}-oriented crystals (compare green curves in Figure \ref{fig:lt_bor} with those in Figure \ref{fig:cor}g). The kinetics consist of three phases. First, the signal decays very rapidly with a slope n of about -4 in the double logarithmic plot. We think that this decay reflects the instrument response function and the time resolution of about 100 ps of our setup. We attribute this signature to the emission of the originally excited singlet exciton, which consists of the 564 and 600 nm band, and its fast decay caused by triplet fission. The primary fission event takes according to literature some picoseconds \cite{Ma2012, Wolf2021} and cannot be resolved with the applied setup.

After three nanoseconds the decay of the emission signal becomes slower but follows again a power law. As in the case of the \textit{c}-oriented crystals, we attribute it to geminate fusion of the triplet excitons formed by the fission event \cite{Wolf2021}. At around 20 ns the decay curve bends and does then not follow any more a power law with a fractional exponent. The signal behaviour in this third phase arises from non-geminate fusion due to migration and diffusion of the triplet excitons \cite{Ryasnyanskiy2011}.

The following function, comprising of three distinct parts, is fitted to the kinetic traces of both sectors for times longer than 2 ns in the case of $\lambda >$ 600 nm and longer than 3 ns in the case of $\lambda <$ 600 nm: 

\begin{equation}
\text{Signal} = A_0e^{-t/\tau} + A_1t^{n} + A_2(\frac{1}{1+T_0\gamma})^{2} + \text{offset}  
\label{eq2}
\end{equation}

The results of the individual fits are given in Table \ref{tabel} and the obtained fit curves are shown in Figure \ref{fig:lt_bor}a and \ref{fig:lt_bor}b as black solid and broken lines. The first term of Equation \ref{eq2} accounts for the exponentially decaying 646 nm band and yields its lifetime of $\tau$ = 3.7 ns.

The second term models by a power-law the emission resulting from geminate fusion of the triplet excitons and their fusion to short-lived but emitting singlet excitons. The obtained power-law exponent is n = -1.5 for both the triangular and diamond zones. The exponent suggests three-dimensional diffusion, similar to the \textit{c}-oriented crystals.

For the third contribution, describing the time dependent of the non-geminate fusion, we obtained $T_0 \gamma \sim 1.9  \mu s^{-1}$ in triangular and $T_0 \gamma \sim 1.3  \mu s^{-1}$, in the diamond zones. By enhancing the excitation power, these evolve up to 10 $\mu s^{-1}$ (see supplementary information). 

These values are significantly smaller in diamond sectors than those for the triangular sectors. This indicates that the non-geminate process depends on the sector type and points to longer migration pathways of the triplet excitons in the diamond sectors. Another contributing factor is probably that contrary to the triangles in the diamond sectors the \textit{c}-axis is turned somewhat out of the plane resulting in a reduced absorption coefficient. This leads to a smaller initial exciton density compared to the triangular sectors and a smaller $T_0\gamma$ value.

\vspace{1em}
\begin{table*}
\centering
\caption{The fitting parameters of the equations (1) and (2).}
\vspace{0.5em} 
\begin{tabular}{c|ccc}
\toprule
orientation & fast & intermediate & slow \\
 & (mono-exp.) & (power law scaling) & (random walk) \\
\midrule
c-oriented diamond    & NA        & n=-1.48        & $T_0 \gamma = 0.61  \mu s^{-1}$        \\
c-oriented triangle    & NA        & n=-1.52        & $T_0 \gamma = 0.62  \mu s^{-1}$        \\
b-oriented diamond ($\lambda <$ 600)    & NA        & n=-1.5        & $T_0 \gamma = 1.25  \mu s^{-1}$        \\
b-oriented diamond ($\lambda >$ 600)    & $\tau$ = 3.8 ns        & n=-1.52        & $T_0 \gamma = 1.24  \mu s^{-1}$        \\
b-oriented triangle ($\lambda <$ 600)    & NA        & n= -1.51        & $T_0 \gamma = 1.88  \mu s^{-1}$        \\
b-oriented triangle ($\lambda >$ 600)   & $\tau$ = 3.6 ns        & n=-1.51        & $T_0 \gamma = 1.92  \mu s^{-1}$        \\
\bottomrule
\end{tabular}
\label{tabel}
\end{table*}

Finally, we discuss shortly the origin of the luminescence band at 646 nm. Strong red-shifted emission around 650 nm has been reported and discussed earlier for amorphous rubrene \cite{Irkhin2012, Chen2011}. Furthermore, a broad band at 650 nm was also observed after melting and cooling of crystalline rubrene \cite{Bardeen2017}. In principle this might also point to an amorphous origin. However, our polarisation sensitive emission measurements show that this luminescence band is very strongly \textit{c}-polarised. This indicates that an amorphous origin is unlikely. As a further explanation for the photoluminescence band at 646 nm of rubrene single crystals, trap states, e.g. resulting from surface oxidation, were discussed \cite{Ma2012, Ma2013}. As described above, we find an exponential decay of this band with a time constant of $\sim$ 3.7 ns. This would be in line with the assumption of a trap state if the lifetime of this state coincides with the observed time constant. On the other hand, we observe after the exponential decay a time evolution of the emission which is identical to that of the two other bands and which reflects the non-geminate fusion dynamics of the triplet excitons. This could hardly be explained by a luminescence due to a trap state.

Here we suggest an alternative explanation. E. Wolf et al. studied \cite{Wolf2018} the photoluminescence kinetics in the presence of a magnetic field and observed a beating signal characteristic for a coherent triplet pair state populated by singlet exciton fission. In addition, they found an exponential decay of the luminescence with a time constant of 4 ns. They assigned it to the lifetime of the triplet pair state. Interestingly, they observed this exponential decay also in absence of a magnetic field. Since their value of 4 ns proposed for the lifetime of the triplet pair state fits very well to the decay time of 3.7 ns for the 646 nm band, we tentatively interpret this spectral feature as luminescence resulting from a triplet pair state. This triplet pair state is formed by fission within a few picoseconds from a singlet exciton and exhibits a total spin of zero \cite{Wolf2018}. Accordingly, a radiative transition to the ground state is spin allowed. The red-shift of the emission band could originate from Herzberg-Teller coupling, where the entanglement of singlet and triplet pair states leads to an avoided crossing and in turn to a joint vibrational potential with a lowered transition energy \cite{Yong2017}.
In this scenario, the photo-induced dynamics can be described as follows. The singlet exciton generated by the optical excitation transforms by the fission step into a triplet pair state which lives for 3.8 ns until it splits into two uncorrelated triplet excitons. Those diffuse by hopping steps through the crystal. At the beginning geminate and later non-geminate fusion can happen which lead again to population of the triplet pair state which is in equilibrium with the singlet exciton state. This results in delayed singlet luminescence as well as delayed emission from the triplet pair state. The fact that the 646 nm band is only occasionally observed could be related to the fact that this emission is strongly \textit{c}-polarised and only dominates when the \textit{c}-axis is parallel to the detected polarisation, as the comparison between the triangular and diamond sectors shows (see Figure \ref{fig:spec_bor}). However, in most studies, rubrene single crystals are investigated which are \textit{c}-oriented, i.e. the \textit{c}-axis is vertical to the surface. In these cases, the 646 nm band is strongly suppressed.

Considering the band structure, rubrene single crystals exhibit substantial dispersion along the $\Gamma-Y$ direction in the Brillouin zone \cite{bredas2012, ishii2010, koch2023, sakamoto2019}. The HOMO-LUMO bandgap thereby varies by $\approx$ 400 mV, leading to broadened density of states of those bands. The strongest emission is expected from the energetically lowest states of the LUMO band to the highest of the HOMO band, compatible with a red-shift. Since the a-orientation usually is along the interface, the perfect alignment of the transition dipole moment with the polarisation of light emitted along b can merely be accomplished with b-oriented crystals. This is the direction where the dispersion makes a difference.

\section{Conclusion}

Novel types of rubrene microcrystals were grown on SiO$_2$ surfaces using superfine powder deposition in an enhanced partial pressure environment, followed by high-rate heating treatment. These crystals are classified into two types based on their growth direction, which determine characteristics of their photoluminescence spectra and kinetics. They exhibit hourglass-shaped zones (of diamond and triangular shape) with distinct optical properties. The origin of this zoned sector growth, which is not reflected in morphology, may be due to anisotropic growth speeds and accordingly a slight rotation from the respective out-of-plane axis in each crystal type. We attribute the long-wavelength band ($\sim646 nm$) is associated with high luminescence intensity with polarisation aligned along c crystal direction. The detected photons originate either from direct emission of geminate coherent triplet pairs or upon fusion of it, exhibiting pure mono-exponential dynamics with 3.7 ns lifetime.

The intermediate time regime appears to be dominated by geminate triplet fusion, however after minimal spatial hopping. The longer time regime is dominated by non-geminate fusion after substantial migration. These crystals hold promise as a standard for further studies on exciton dynamics and for revealing the internal properties of rubrene crystalline structures.

\section{Acknowledgement}
Funding by the Deutsche Forschungsgemeinschaft (DFG, German Research Foundation) within projects SFB 1477 ”Light-Matter Interactions at Interfaces” (441234705), SFB 1270/2 ”Electrically Active Implants” (299150580), “Application of Interoperable Metadata Standards (AIMS) 2” (432233186), and for the transmission electron microscope Jeol JEM-ARM200F NeoARM STEM (DFG INST 264/161-1 FUGG) is acknowledged.

\bibliographystyle{unsrt}
\bibliography{ref}

\end{document}